\documentclass[letterpaper,compsoc,twoside]{IEEEtran}
\usepackage{fixltx2e} 
\usepackage{cmap} 
\usepackage{ifthen}
\usepackage[T1]{fontenc}
\usepackage[utf8]{inputenc}
\usepackage{amsmath}

\usepackage[font={small,it},labelfont=bf]{caption}
\usepackage{float}

\usepackage{longtable,ltcaption,array}
\newlength{\DUtablewidth} 
\usepackage{textcomp} 

\pdfoutput=1
\usepackage{scipy}
\makeatletter
\def\PY@reset{\let\PY@it=\relax \let\PY@bf=\relax%
    \let\PY@ul=\relax \let\PY@tc=\relax%
    \let\PY@bc=\relax \let\PY@ff=\relax}
\def\PY@tok#1{\csname PY@tok@#1\endcsname}
\def\PY@toks#1+{\ifx\relax#1\empty\else%
    \PY@tok{#1}\expandafter\PY@toks\fi}
\def\PY@do#1{\PY@bc{\PY@tc{\PY@ul{%
    \PY@it{\PY@bf{\PY@ff{#1}}}}}}}
\def\PY#1#2{\PY@reset\PY@toks#1+\relax+\PY@do{#2}}

\expandafter\def\csname PY@tok@gd\endcsname{\def\PY@tc##1{\textcolor[rgb]{0.63,0.00,0.00}{##1}}}
\expandafter\def\csname PY@tok@gu\endcsname{\let\PY@bf=\textbf\def\PY@tc##1{\textcolor[rgb]{0.50,0.00,0.50}{##1}}}
\expandafter\def\csname PY@tok@gt\endcsname{\def\PY@tc##1{\textcolor[rgb]{0.00,0.27,0.87}{##1}}}
\expandafter\def\csname PY@tok@gs\endcsname{\let\PY@bf=\textbf}
\expandafter\def\csname PY@tok@gr\endcsname{\def\PY@tc##1{\textcolor[rgb]{1.00,0.00,0.00}{##1}}}
\expandafter\def\csname PY@tok@cm\endcsname{\let\PY@it=\textit\def\PY@tc##1{\textcolor[rgb]{0.25,0.50,0.56}{##1}}}
\expandafter\def\csname PY@tok@vg\endcsname{\def\PY@tc##1{\textcolor[rgb]{0.73,0.38,0.84}{##1}}}
\expandafter\def\csname PY@tok@m\endcsname{\def\PY@tc##1{\textcolor[rgb]{0.13,0.50,0.31}{##1}}}
\expandafter\def\csname PY@tok@mh\endcsname{\def\PY@tc##1{\textcolor[rgb]{0.13,0.50,0.31}{##1}}}
\expandafter\def\csname PY@tok@cs\endcsname{\def\PY@tc##1{\textcolor[rgb]{0.25,0.50,0.56}{##1}}\def\PY@bc##1{\setlength{\fboxsep}{0pt}\colorbox[rgb]{1.00,0.94,0.94}{\strut ##1}}}
\expandafter\def\csname PY@tok@ge\endcsname{\let\PY@it=\textit}
\expandafter\def\csname PY@tok@vc\endcsname{\def\PY@tc##1{\textcolor[rgb]{0.73,0.38,0.84}{##1}}}
\expandafter\def\csname PY@tok@il\endcsname{\def\PY@tc##1{\textcolor[rgb]{0.13,0.50,0.31}{##1}}}
\expandafter\def\csname PY@tok@go\endcsname{\def\PY@tc##1{\textcolor[rgb]{0.20,0.20,0.20}{##1}}}
\expandafter\def\csname PY@tok@cp\endcsname{\def\PY@tc##1{\textcolor[rgb]{0.00,0.44,0.13}{##1}}}
\expandafter\def\csname PY@tok@gi\endcsname{\def\PY@tc##1{\textcolor[rgb]{0.00,0.63,0.00}{##1}}}
\expandafter\def\csname PY@tok@gh\endcsname{\let\PY@bf=\textbf\def\PY@tc##1{\textcolor[rgb]{0.00,0.00,0.50}{##1}}}
\expandafter\def\csname PY@tok@ni\endcsname{\let\PY@bf=\textbf\def\PY@tc##1{\textcolor[rgb]{0.84,0.33,0.22}{##1}}}
\expandafter\def\csname PY@tok@nl\endcsname{\let\PY@bf=\textbf\def\PY@tc##1{\textcolor[rgb]{0.00,0.13,0.44}{##1}}}
\expandafter\def\csname PY@tok@nn\endcsname{\let\PY@bf=\textbf\def\PY@tc##1{\textcolor[rgb]{0.05,0.52,0.71}{##1}}}
\expandafter\def\csname PY@tok@no\endcsname{\def\PY@tc##1{\textcolor[rgb]{0.38,0.68,0.84}{##1}}}
\expandafter\def\csname PY@tok@na\endcsname{\def\PY@tc##1{\textcolor[rgb]{0.25,0.44,0.63}{##1}}}
\expandafter\def\csname PY@tok@nb\endcsname{\def\PY@tc##1{\textcolor[rgb]{0.00,0.44,0.13}{##1}}}
\expandafter\def\csname PY@tok@nc\endcsname{\let\PY@bf=\textbf\def\PY@tc##1{\textcolor[rgb]{0.05,0.52,0.71}{##1}}}
\expandafter\def\csname PY@tok@nd\endcsname{\let\PY@bf=\textbf\def\PY@tc##1{\textcolor[rgb]{0.33,0.33,0.33}{##1}}}
\expandafter\def\csname PY@tok@ne\endcsname{\def\PY@tc##1{\textcolor[rgb]{0.00,0.44,0.13}{##1}}}
\expandafter\def\csname PY@tok@nf\endcsname{\def\PY@tc##1{\textcolor[rgb]{0.02,0.16,0.49}{##1}}}
\expandafter\def\csname PY@tok@si\endcsname{\let\PY@it=\textit\def\PY@tc##1{\textcolor[rgb]{0.44,0.63,0.82}{##1}}}
\expandafter\def\csname PY@tok@s2\endcsname{\def\PY@tc##1{\textcolor[rgb]{0.25,0.44,0.63}{##1}}}
\expandafter\def\csname PY@tok@vi\endcsname{\def\PY@tc##1{\textcolor[rgb]{0.73,0.38,0.84}{##1}}}
\expandafter\def\csname PY@tok@nt\endcsname{\let\PY@bf=\textbf\def\PY@tc##1{\textcolor[rgb]{0.02,0.16,0.45}{##1}}}
\expandafter\def\csname PY@tok@nv\endcsname{\def\PY@tc##1{\textcolor[rgb]{0.73,0.38,0.84}{##1}}}
\expandafter\def\csname PY@tok@s1\endcsname{\def\PY@tc##1{\textcolor[rgb]{0.25,0.44,0.63}{##1}}}
\expandafter\def\csname PY@tok@gp\endcsname{\let\PY@bf=\textbf\def\PY@tc##1{\textcolor[rgb]{0.78,0.36,0.04}{##1}}}
\expandafter\def\csname PY@tok@sh\endcsname{\def\PY@tc##1{\textcolor[rgb]{0.25,0.44,0.63}{##1}}}
\expandafter\def\csname PY@tok@ow\endcsname{\let\PY@bf=\textbf\def\PY@tc##1{\textcolor[rgb]{0.00,0.44,0.13}{##1}}}
\expandafter\def\csname PY@tok@sx\endcsname{\def\PY@tc##1{\textcolor[rgb]{0.78,0.36,0.04}{##1}}}
\expandafter\def\csname PY@tok@bp\endcsname{\def\PY@tc##1{\textcolor[rgb]{0.00,0.44,0.13}{##1}}}
\expandafter\def\csname PY@tok@c1\endcsname{\let\PY@it=\textit\def\PY@tc##1{\textcolor[rgb]{0.25,0.50,0.56}{##1}}}
\expandafter\def\csname PY@tok@kc\endcsname{\let\PY@bf=\textbf\def\PY@tc##1{\textcolor[rgb]{0.00,0.44,0.13}{##1}}}
\expandafter\def\csname PY@tok@c\endcsname{\let\PY@it=\textit\def\PY@tc##1{\textcolor[rgb]{0.25,0.50,0.56}{##1}}}
\expandafter\def\csname PY@tok@mf\endcsname{\def\PY@tc##1{\textcolor[rgb]{0.13,0.50,0.31}{##1}}}
\expandafter\def\csname PY@tok@err\endcsname{\def\PY@bc##1{\setlength{\fboxsep}{0pt}\fcolorbox[rgb]{1.00,0.00,0.00}{1,1,1}{\strut ##1}}}
\expandafter\def\csname PY@tok@kd\endcsname{\let\PY@bf=\textbf\def\PY@tc##1{\textcolor[rgb]{0.00,0.44,0.13}{##1}}}
\expandafter\def\csname PY@tok@ss\endcsname{\def\PY@tc##1{\textcolor[rgb]{0.32,0.47,0.09}{##1}}}
\expandafter\def\csname PY@tok@sr\endcsname{\def\PY@tc##1{\textcolor[rgb]{0.14,0.33,0.53}{##1}}}
\expandafter\def\csname PY@tok@mo\endcsname{\def\PY@tc##1{\textcolor[rgb]{0.13,0.50,0.31}{##1}}}
\expandafter\def\csname PY@tok@mi\endcsname{\def\PY@tc##1{\textcolor[rgb]{0.13,0.50,0.31}{##1}}}
\expandafter\def\csname PY@tok@kn\endcsname{\let\PY@bf=\textbf\def\PY@tc##1{\textcolor[rgb]{0.00,0.44,0.13}{##1}}}
\expandafter\def\csname PY@tok@o\endcsname{\def\PY@tc##1{\textcolor[rgb]{0.40,0.40,0.40}{##1}}}
\expandafter\def\csname PY@tok@kr\endcsname{\let\PY@bf=\textbf\def\PY@tc##1{\textcolor[rgb]{0.00,0.44,0.13}{##1}}}
\expandafter\def\csname PY@tok@s\endcsname{\def\PY@tc##1{\textcolor[rgb]{0.25,0.44,0.63}{##1}}}
\expandafter\def\csname PY@tok@kp\endcsname{\def\PY@tc##1{\textcolor[rgb]{0.00,0.44,0.13}{##1}}}
\expandafter\def\csname PY@tok@w\endcsname{\def\PY@tc##1{\textcolor[rgb]{0.73,0.73,0.73}{##1}}}
\expandafter\def\csname PY@tok@kt\endcsname{\def\PY@tc##1{\textcolor[rgb]{0.56,0.13,0.00}{##1}}}
\expandafter\def\csname PY@tok@sc\endcsname{\def\PY@tc##1{\textcolor[rgb]{0.25,0.44,0.63}{##1}}}
\expandafter\def\csname PY@tok@sb\endcsname{\def\PY@tc##1{\textcolor[rgb]{0.25,0.44,0.63}{##1}}}
\expandafter\def\csname PY@tok@k\endcsname{\let\PY@bf=\textbf\def\PY@tc##1{\textcolor[rgb]{0.00,0.44,0.13}{##1}}}
\expandafter\def\csname PY@tok@se\endcsname{\let\PY@bf=\textbf\def\PY@tc##1{\textcolor[rgb]{0.25,0.44,0.63}{##1}}}
\expandafter\def\csname PY@tok@sd\endcsname{\let\PY@it=\textit\def\PY@tc##1{\textcolor[rgb]{0.25,0.44,0.63}{##1}}}

\def\PYZus{\char`\_}

\def\PYZsh{\char`\#}

\def\PYZhy{\char`\-}
\def\PYZsq{\char`\'}

\makeatother

\providecommand*{\DUfootnotemark}[3]{%
  \raisebox{1em}{\hypertarget{#1}{}}%
  \hyperlink{#2}{\textsuperscript{#3}}%
}
\providecommand{\DUfootnotetext}[4]{%
  \begingroup%
  \renewcommand{\thefootnote}{%
    \protect\raisebox{1em}{\protect\hypertarget{#1}{}}%
    \protect\hyperlink{#2}{#3}}%
  \footnotetext{#4}%
  \endgroup%
}

\providecommand*{\DUrole}[2]{%
  \ifcsname DUrole#1\endcsname%
    \csname DUrole#1\endcsname{#2}%
  \else
    \ifcsname docutilsrole#1\endcsname%
      \csname docutilsrole#1\endcsname{#2}%
    \else%
      #2%
    \fi%
  \fi%
}

\ifthenelse{\isundefined{\hypersetup}}{
  \usepackage[colorlinks=true,linkcolor=blue,urlcolor=blue]{hyperref}
  \urlstyle{same} 
}{}

\begin{document}
\newcounter{footnotecounter}\title{Computing an Optimal Control Policy for an Energy Storage}\author{Pierre Haessig$^{\setcounter{footnotecounter}{1}\fnsymbol{footnotecounter}\setcounter{footnotecounter}{2}\fnsymbol{footnotecounter}}$%
          \setcounter{footnotecounter}{1}\thanks{\fnsymbol{footnotecounter} %
          Corresponding author: \protect\href{mailto:pierre.haessig@ens-rennes.fr}{pierre.haessig@ens-rennes.fr}}\setcounter{footnotecounter}{2}\thanks{\fnsymbol{footnotecounter} SATIE CNRS laboratory - ENS Rennes, Bruz, France}, Thibaut Kovaltchouk$^{\setcounter{footnotecounter}{2}\fnsymbol{footnotecounter}}$, Bernard Multon$^{\setcounter{footnotecounter}{2}\fnsymbol{footnotecounter}}$, Hamid Ben Ahmed$^{\setcounter{footnotecounter}{2}\fnsymbol{footnotecounter}}$, Stéphane Lascaud$^{\setcounter{footnotecounter}{3}\fnsymbol{footnotecounter}}$\setcounter{footnotecounter}{3}\thanks{\fnsymbol{footnotecounter} LME department - EDF R\&D, Écuelles, France}\thanks{%

          \noindent%
          Copyright\,\copyright\,2014 Pierre Haessig et al. This is an open-access article distributed under the terms of the Creative Commons Attribution License, which permits unrestricted use, distribution, and reproduction in any medium, provided the original author and source are credited. http://creativecommons.org/licenses/by/3.0/%
        }}\maketitle
          \renewcommand{\leftmark}{PROC. OF THE 6th EUR. CONF. ON PYTHON IN SCIENCE (EUROSCIPY 2013)}
          \renewcommand{\rightmark}{COMPUTING AN OPTIMAL CONTROL POLICY FOR AN ENERGY STORAGE}

\setcounter{page}{51}
\newcommand*{\docutilsroleref}{\ref}
\newcommand*{\docutilsrolelabel}{\label}
\AtEndDocument{\cleardoublepage}
\begin{abstract}We introduce StoDynProg, a small library created to solve Optimal
Control problems arising in the management of Renewable Power Sources,
in particular when coupled with an Energy Storage System. The library
implements generic Stochastic Dynamic Programming (SDP) numerical
methods which can solve a large class of Dynamic Optimization problems.

We demonstrate the library capabilities with a prototype problem:
smoothing the power of an Ocean Wave Energy Converter. First we use time
series analysis to derive a stochastic Markovian model of this system
since it is required by Dynamic Programming. Then, we briefly describe
the “policy iteration” algorithm we have implemented and the numerical
tools being used. We show how the API design of the library is generic
enough to address Dynamic Optimization problems outside the field of
Energy Management. Finally, we solve the power smoothing problem and
compare the optimal control with a simpler heuristic control.\end{abstract}

\section{Introduction to Power Production Smoothing%
  \label{introduction-to-power-production-smoothing}%
  \label{s-intro-smoothing}%
}

Electric power generated by renewable
sources like wind, sun or ocean waves can exhibit a strong \emph{variability}
along time. Because on an electricity grid the energy production must
match the consumption, this variability can be an issue for the grid
stability. Yet most of the time, fluctuations of renewable power sources
are absorbed without trouble thanks to regulation mechanisms which make
flexible generation units adjust their production in real-time.
Therefore, the production-consumption equilibrium can be maintained.

However, there are cases where fluctuations may be considered too strong
to be fed directly to the grid so that an \emph{energy storage system},
acting as a \emph{buffer}, may be required to smooth out the production. The
schematic of the system considered in this article is given on figure
\DUrole{ref}{smoothing-diagram}.\begin{figure}[t]\noindent\makebox[\columnwidth][c]{\includegraphics[width=\columnwidth]{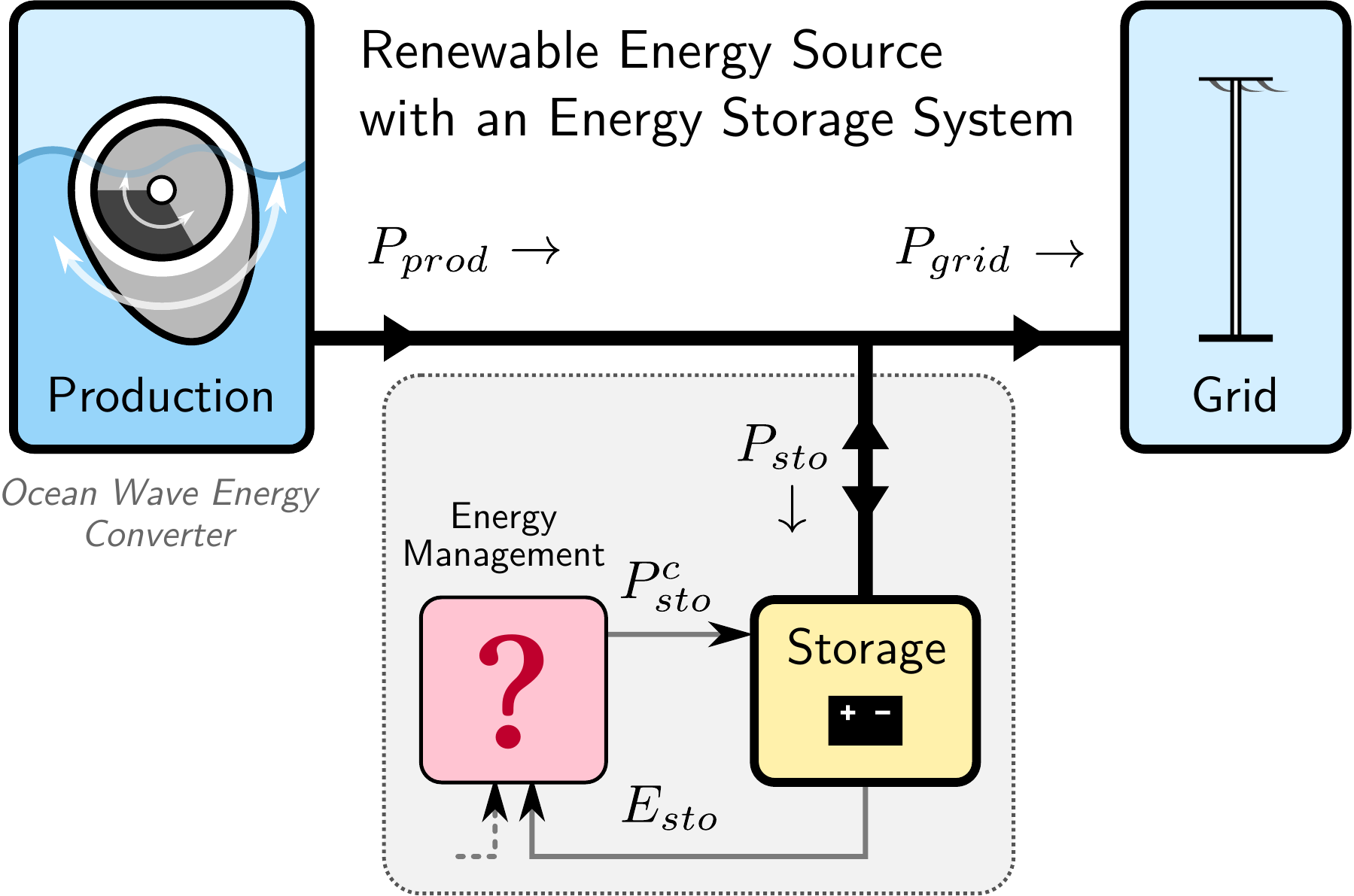}}
\caption{Power smoothing with an Energy Storage: an example of an Optimal Control problem.
\DUrole{label}{smoothing-diagram}}
\end{figure}

\subsection{Smoothing with an Energy Storage%
  \label{smoothing-with-an-energy-storage}%
}

Electricity generation from \emph{ocean waves} (with machines called Wave
Energy Converters) is an example where the output power can be \emph{strongly
fluctuating}. This is illustrated on figure \DUrole{ref}{smooth-lin}
where the output power $P_{prod}(t)$ from a particular wave energy
converter called SEAREV is represented during 100 seconds.

We just mention that this production time series does not come from
measurements but from an hydro-mechanical simulation from colleagues
since the SEAREV is a big 1 MW - 30 meters long machine which is yet to be
built \cite{Ruellan-2010}.

The oscillations of $P_{prod}(t)$ at a period of about 1.5 s comes
from the construction of the SEAREV: in short, it is a floating
\emph{double-pendulum} that oscillates with the waves. Also, because \emph{ocean
waves have a stochastic behavior}, the amplitude of these oscillations
is irregular.\begin{figure}[]\noindent\makebox[\columnwidth][c]{\includegraphics[width=\columnwidth]{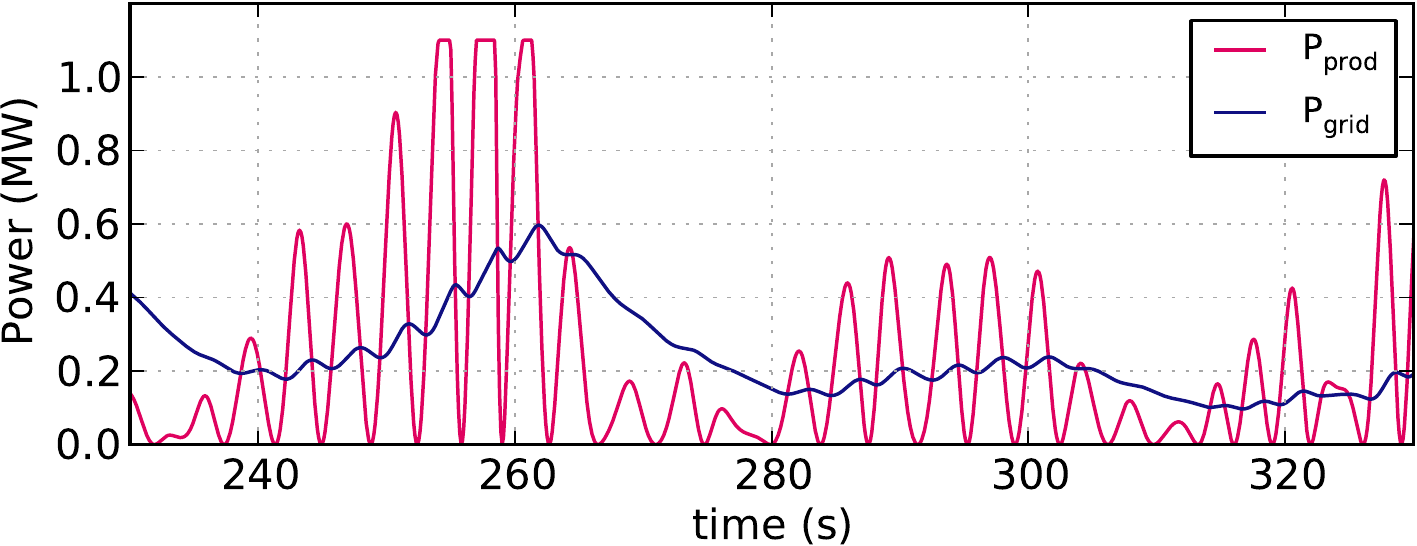}}
\caption{Smoothing the Ocean Power injected to the grid using an Energy Storage
controlled by the simple linear law (\DUrole{ref}{eq-feedback-lin}).
The storage buffers the difference between the two powers.
\DUrole{label}{smooth-lin}}
\end{figure}

Therefore an energy storage absorbing a power $P_{sto}$ can be
used to smooth out the power $P_{grid}$ injected to the
electricity network:\begin{equation}
\label{eq-P-grid}
 P_{grid}(t) = P_{prod}(t) - P_{sto}(t)
\end{equation}The energy of the storage then evolves as:\begin{equation}
\label{eq-E-sto}
 E_{sto}(k+1) = E_{sto}(k) + P_{sto}(k)\Delta t
\end{equation}expressed here in discrete time ($\Delta t = 0.1 \text{ s}$
throughout this article), without accounting for losses. The storage
energy is bounded: $0 \leq E_{sto} \leq E_{rated}$, where
$E_{rated}$ denotes the storage capacity which is set to 10 MJ in
this article (i.e. about 10 seconds of reserve at full power)

It is a control problem to choose a power smoothing law. We present the
example of a linear feedback control:\begin{equation}
\label{eq-feedback-lin}
 P_{grid}(t) = \frac{P_{max}}{E_{rated}} E_{sto}(t)
\end{equation}where $P_{max}$ is the rated power of the SEAREV (1.1 MW). This law gives
“good enough” smoothing results as it can be seen on figure
\DUrole{ref}{smooth-lin}.

The performance of the smoothing is greatly influenced by the \emph{storage
sizing} (i.e. the choice of the capacity $E_{rated}$). This
question is not addressed in this article but was discussed by
colleagues \cite{Aubry-2010}. We also don’t discuss the choice of the
storage \emph{technology}, but it is believed that super-capacitors would be
the most suitable choice.
Because energy storage is very expensive (\textasciitilde{}20 k€/kWh or \textasciitilde{}5 k€/MJ for supercaps),
there is an interest in studying how to make the best use of a
given capacity to avoid a costly over-sizing.

\subsection{Finding an Optimal Smoothing Policy%
  \label{finding-an-optimal-smoothing-policy}%
}

Control law (\DUrole{ref}{eq-feedback-lin}) is an example of heuristic choice
of policy and we now try
to go further by finding an \emph{optimal} policy.

Optimality will be measured against a \emph{cost function} $J$ that
penalizes the average variability of the power injected to the grid:\begin{equation}
\label{eq-cost}
J = \frac{1}{N} \mathbb{E} \left\lbrace
                   \sum_{k=0}^{N-1} c(P_{grid}(k))
                 \right\rbrace
                 \quad
                 \text{with $N \rightarrow \infty$}
\end{equation}where $c$ is the \emph{instantaneous cost} (or penalty) function which
can be $c(P_{grid}) = P_{grid}^2$ for example.
Expectation~$\mathbb{E}$ is needed because the production
$P_{prod}$ is a stochastic input, so that the output power
$P_{grid}$ is also a random variable.

This minimization problem falls in the class of \emph{stochastic dynamic
optimization}. It is \emph{dynamic} because decisions at each time-step
cannot be taken independently due to the coupling along time introduced by the
evolution of the stored energy (\DUrole{ref}{eq-E-sto}). To describe the dynamics of the system,
we use the generic notation\begin{equation}
\label{eq-state-dyn}
x_{k+1} = f(x_{k}, u_k, \varepsilon_k)
\end{equation}where $x, u, \varepsilon$ are respectively \emph{state} variables,
\emph{control} variables and \emph{perturbations}. State variables are the
“memory” of the system. The stored energy $E_{sto}$ is here the
only state variable, but more will appear in section \DUrole{ref}{ss-ss-model}.
Control variables are the ones which values must be chosen at each instant
to optimize the cost $J$.
The injected power $P_{grid}$ is here the single control variable.

Dynamic optimization (also called \emph{optimal control}) is addressed by the
Dynamic Programming method \cite{Bertsekas-2005} which yields a theoretical
analysis of the solution structure. Indeed, once all state variables
(i.e. “memories”) of the system are identified, the optimum of the cost
$J$ is attained by a “state feedback” policy, that is a policy
where the control is chosen as \emph{a function of the state:}\begin{equation}
\label{eq-feedback-opt}
 P_{grid}(t) = \mu(x(t))
\end{equation}The goal is then to find the \emph{optimal} feedback function $\mu$.
Since $E_{sto}$ is a state variable, policy (\DUrole{ref}{eq-feedback-lin})
is in fact a special case of (\DUrole{ref}{eq-feedback-opt}).
Since $\mu$ has no special structure in the general
case\DUfootnotemark{id4}{id13}{1}, it will be \emph{numerically computed on a grid} over the state
space. We cover the algorithm for this computation in
section~\DUrole{ref}{s-opt-sto-ctrl}.

\subsubsection{Prerequisite%
  \label{prerequisite}%
}

Dynamic Programming does require that stochastic perturbations are
\emph{independent} random variables (i.e. the overall dynamical model must be
Markovian) and this is not true for the $P_{prod}(k)$ time series.
Therefore we devote section~\DUrole{ref}{s-stoch-model} to the problem of
expressing $P_{prod}$ as a discrete-time Markov process, using
\emph{time series analysis}. This will yield new state variables accounting
for the dynamics of $P_{prod}$.

\section{Stochastic Model of a Wave Energy Production%
  \label{stochastic-model-of-a-wave-energy-production}%
  \label{s-stoch-model}%
}

We now take a closer look at the $P_{prod}$ time series. A 1000~s
long simulation is presented on figure~\DUrole{ref}{speed-pow}, along
with a zoom to better see the structure at short time scales. An
histogram is also provided which shows that $P_{prod}$ is clearly
\emph{non-gaussian}. This precludes the direct use of “standard” time series
models based on Autoregressive Moving Average (ARMA) models \cite{Brockwell-1991}.

However, we can leverage the knowledge of the inner working of the SEAREV.
Indeed, by calling $\Omega$ the rotational speed of the inner
pendulum with respect to the hull, we know that the output power is:\begin{equation}
\label{eq-P-prod}
 P_{prod} = T_{PTO}(\Omega).\Omega
\end{equation}where $T_{PTO}$ is the torque applied to the pendulum by the
electric machine which harvests the energy (PTO stands for “Power Take
Off”). Finding the best $T_{PTO}$ command is actually another
optimal control problem which is still an active area of research in the
Wave Energy Conversion community \cite{Kovaltchouk-2013}. We use here a
“viscous damping law, with power leveling”, that is
$T_{PTO}(\Omega) = \beta.\Omega$. This law is applied as long as
it yields a power below $P_{max}$. Otherwise the torque is reduced
to level the power at 1.1~MW as can be seen on figure
\DUrole{ref}{speed-pow} whenever the speed is more than 0.5~rad/s.

Thanks to equation (\DUrole{ref}{eq-P-prod}), we can thus model the speed $\Omega$ and then
deduce $P_{prod}$. Modeling the speed is much easier because it is
quite Gaussian (see fig. \DUrole{ref}{speed-pow}) and has a much more
regular behavior which can be captured by an ARMA process.

\subsection{Autoregressive Model of the Speed%
  \label{autoregressive-model-of-the-speed}%
}

Within the ARMA family, we restrict ourselves to the autoregressive
(AR) processes because we need a Markovian model. The equation of an AR(p)
model for the speed is:\begin{equation}
\label{eq-ar-p}
\Omega(k) = \phi_1 \Omega(k-1) + \dots +  \phi_p \Omega(k-p) +  \varepsilon(k)
\end{equation}where $p$ is the order of the model and $\varepsilon(k)$ is
a series independent random variables. Equation (\DUrole{ref}{eq-ar-p}) indeed yields a
Markovian process, using the lagged observations of the speed
$\Omega(k-1), \dots, \Omega(k-p)$ as state variables.

AR(p) model fitting consists in \emph{selecting} the order $p$ and
\emph{estimating} the unknown coefficients $\phi_1, \dots, \phi_p$ as
well as the unknown variance of $\varepsilon$ which we denote
$\sigma_{\varepsilon}^2$.\begin{figure}[]\noindent\makebox[\columnwidth][c]{\includegraphics[width=\columnwidth]{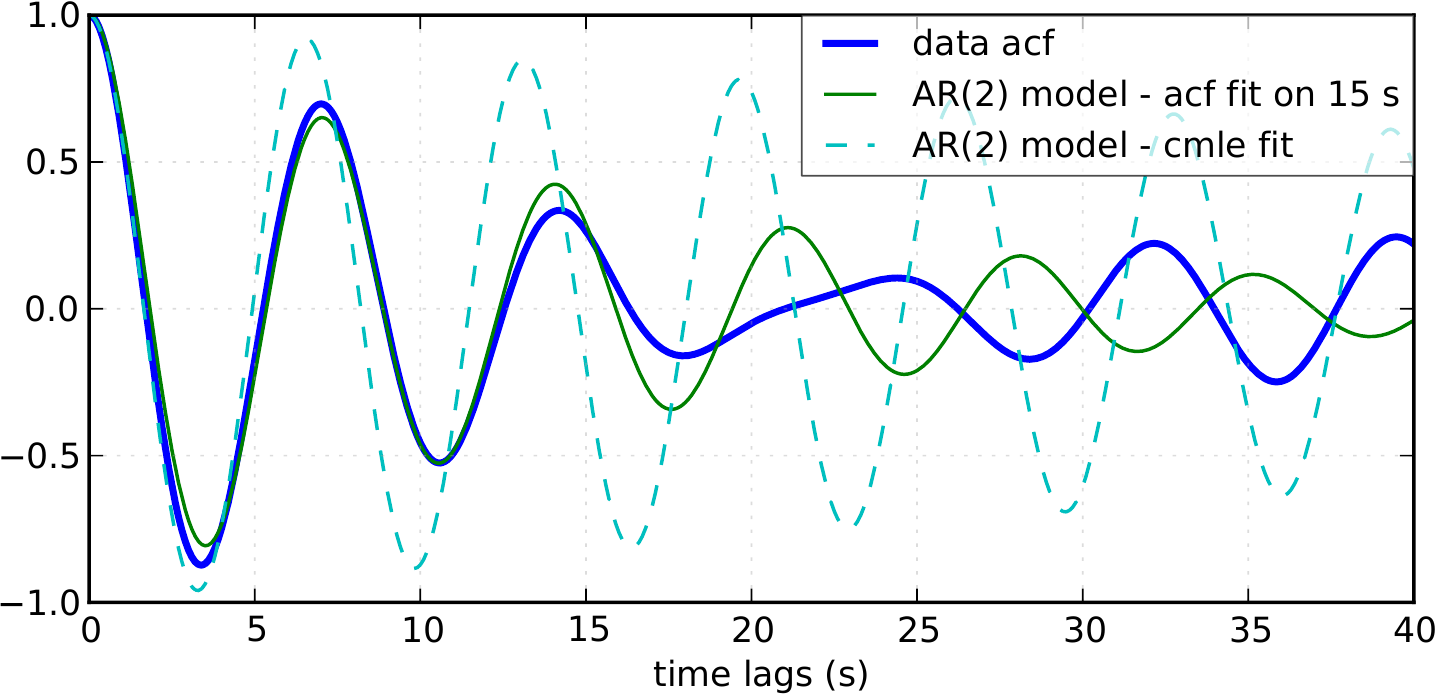}}
\caption{Autocorrelation function (acf) of the speed data,
compared with the acf from two AR(2) models,
fitted with two different methods.
\DUrole{label}{fig-speed-acf-AR2}}
\end{figure}\begin{figure*}[]\noindent\makebox[\textwidth][c]{\includegraphics[scale=0.70]{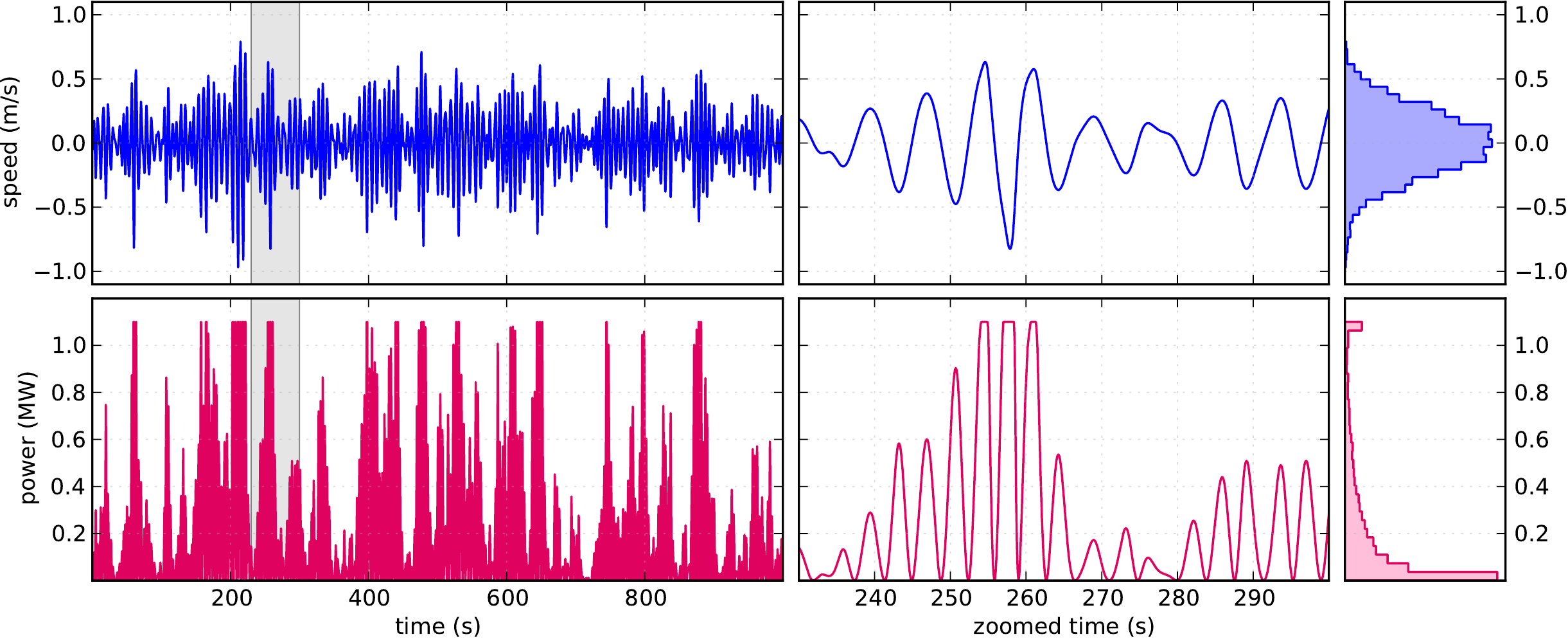}}
\caption{Speed \& Power time series from a 1000 seconds SEAREV simulation (sample Em\_1.txt).
The gray rectangle time interval is enlarged in the middle panel.
Distribution histogram on the right.
\DUrole{label}{speed-pow}}
\end{figure*}

\subsubsection{Order selection%
  \label{order-selection}%
}

is generally done using \emph{information criterions} such as AIC or
BIC \cite{Brockwell-1991}, but for this modeling problem, we
restrict ourselves to the smallest order which can reproduce the
\emph{decaying oscillations} of the autocorrelation function.
Autocorrelation (acf) of the speed
is plotted on figure \DUrole{ref}{fig-speed-acf-AR2} where we can
see that a model of order $p=2$ can indeed reproduce the
autocorrelation up to about 15 s of time lags
(a 1\textsuperscript{st} order model would only yield an exponential decay without oscillations).
15 s is thought to be the time horizon of interest when using a 10 MJ/1.1 MW
energy storage.

Keeping the model order low is required to maintain the dimension of the
overall state vector under 3 or 4. The underlying issue of an
exponentially growing complexity will appear in section
\DUrole{ref}{s-opt-sto-ctrl} when solving the Dynamic Programming
equation.

\subsubsection{Parameter estimation%
  \label{parameter-estimation}%
}

Once the order is selected, we have to estimate coefficients
$\phi_1$, $\phi_2$ and $\sigma_{\varepsilon}^2$.
“Classical” fitting methodology \cite{Brockwell-1991} is based on
Conditional Maximum Likelihood Estimators (CMLE). This method is readily
available in \texttt{GNU R} with the \texttt{arima} routine or in Python with
\texttt{statsmodel.tsa.ar\_model}.

However, we have plotted the autocorrelation of the estimated AR(2)
model on figure \DUrole{ref}{fig-speed-acf-AR2} to show that CMLE
is \emph{not appropriate:} oscillations of the acf clearly decay too slowly
compared to the data acf.

The poor adequacy of this fit is actually a consequence of our choice of
a low order model which implies that the AR(2) process can only be an
\emph{approximation of the true process}. Statistically speaking, our model
is \emph{misspecified}, whereas CMLE is efficient for correctly specified
models only. This problem has been discussed in the literature
\cite{McElroy-2013} and has yielded the “Multi-step ahead fitting procedure”.

Being unfamiliar with the latter approach, we compute instead
$\phi_1, \phi_2$ estimates which \emph{minimize the difference} between
the theoretical AR(2) acf and the data acf. The minimization criterion
is the sum of the squared acf differences over a range of lag times
which can be chosen. We name this approach the “multi-lags acf fitting”
method. Minimization is conducted with \texttt{fmin} from \texttt{scipy.optimize}.\begin{table}
\setlength{\DUtablewidth}{0.8\linewidth}
\begin{longtable*}[c]{|p{0.147\DUtablewidth}|p{0.226\DUtablewidth}|p{0.226\DUtablewidth}|p{0.350\DUtablewidth}|}
\hline

method & 

$\hat{\phi}_1$ & 

$\hat{\phi}_2$ & 

$\hat{\sigma}_{\varepsilon}$ \\
\hline

CMLE & 

1.9883 (.0007) & 

-0.9975 (.0007) & 

0.00172 \\
\hline

fit on 15 s & 

1.9799 & 

-0.9879 & 

0.00347 \\
\hline
\end{longtable*}
\caption{AR(2) fitting results from the two methods (along with standard error when available).
\DUrole{label}{tab-ar2-fit}}\end{table}

The result of this acf fitting over time lags up to 15 s (i.e. 150 lags)
is shown on figure~\DUrole{ref}{fig-speed-acf-AR2} while
numerical estimation results are given in table~\DUrole{ref}{tab-ar2-fit}.

With the model obtained from this multi-lags method, we can simulate
speed and power trajectories and check that they have a “realistic
behavior”. We can thus infer that the dynamic optimization algorithm
should make appropriate control decisions out of it. This will be
discussed in section \DUrole{ref}{ss-results-searev-smooth}.
Going further, it would be interesting to study the influence of the AR parameters
(including order $p$)
on the dynamic optimization to see if the “multi-lags acf fitting” indeed brings
an improvement of the final cost function $J$.

\subsection{Reformulation as a state-space model%
  \label{reformulation-as-a-state-space-model}%
  \label{ss-ss-model}%
}

The AR(2) model is a state-space model with state
variables being the lagged observations of the speed $\Omega(k-1)$
and $\Omega(k-2)$. In order to get a model with a better “physical
interpretation” we introduce the variable
$A_k = (\Omega_k - \Omega_{k-1})/\Delta t$ which is the backward
discrete derivative of $\Omega$. As the timestep gets smaller
$A_k$ comes close to the acceleration (in rad/s\textsuperscript{2}) of the
pendulum. Using $(\Omega, A)$ as the state vector, we obtain the
following state-space model:\begin{equation}
\label{eq-ss-ar2}
\begin{split}
  \begin{pmatrix}
   \Omega_k\\
   A_k
  \end{pmatrix}
  =&
  \begin{bmatrix}
   \phi_1 + \phi_2&                -\phi_2 \Delta t\\
   (\phi_1 + \phi_2 - 1)/\Delta t& -\phi_2
  \end{bmatrix}
  \begin{pmatrix}
   \Omega_{k-1}\\
   A_{k-1}
  \end{pmatrix}\\
  +&
  \begin{bmatrix}
   1\\
   1/\Delta t
  \end{bmatrix}
  \varepsilon_k
\end{split}
\end{equation}We now have a stochastic Markovian model for the power production of the SEAREV.
Taken together with state equation of the storage (\DUrole{ref}{eq-E-sto})
and algebraic relations (\DUrole{ref}{eq-P-grid}) and (\DUrole{ref}{eq-P-prod}),
we have a Markovian model of the overall system. The
state vector $x=(E_{sto}, \Omega, A)$ is of dimension 3 which is
just small enough to apply the Stochastic Dynamic Programming method.

\section{Optimal storage control with Dynamic Programming%
  \label{optimal-storage-control-with-dynamic-programming}%
  \label{s-opt-sto-ctrl}%
}

\subsection{The Policy Iteration Algorithm%
  \label{the-policy-iteration-algorithm}%
  \label{ss-pol-iter}%
}

We now give an overview of the \emph{policy iteration}
algorithm that we implemented to solve the power smoothing problem
described in the introduction. Among the different types of dynamic
optimization problems, it is an “infinite horizon, average cost per
stage problem” (as seen in (\DUrole{ref}{eq-cost})). While at first this cost equation involves
a summation over an infinite number of instants, the Dynamic Programming
approach cuts this into two terms: the present and the whole future. In
the end, the optimization falls back to solving this equilibrium
equation:\begin{equation}
\label{eq-dp-avg-equil}
 \begin{split}
 J + \tilde{J}(x) = \min_{u \in U(x)}                \underset{w}{\mathbb{E}}               \Big\lbrace                 \underbrace{ c(x, u, w)
                           }_{\text{instant cost}}\\
            +
            \underbrace{  \tilde{J}(f(x, u, w))
                           }_{\text{cost of the future}}
             \Big\rbrace   \end{split}
\end{equation}where $J$ is the minimized average cost and $\tilde{J}$ is
the transient (or differential) cost function, also called \emph{value
function}.

Note that eq. (\DUrole{ref}{eq-dp-avg-equil}) is a functional equation for $\tilde{J}$ which
should be solved for \emph{any value} of the state $x$ in the state
space. In practice, it is solved in a \emph{discrete grid} that must be
chosen so that the variations of $\tilde{J}$ are represented with
enough accuracy. Also, the optimal policy $\mu$ appears implicitly
as the \emph{argmin} of this equation, that is the optimal control $u$
for each $x$ value of the state grid.

\subsubsection{Equation solving%
  \label{equation-solving}%
}

The simplest way to solve eq. (\DUrole{ref}{eq-dp-avg-equil}) is to iterate the right-hand side,
starting with a zero value function. This is called \emph{value iteration}.

A more efficient approach is \emph{policy iteration}. It starts with an
initial policy (like the heuristic linear (\DUrole{ref}{eq-feedback-lin}))
and gradually improves it with a two steps procedure:\newcounter{listcnt0}
\begin{list}{\arabic{listcnt0}.}
{
\usecounter{listcnt0}
\setlength{\rightmargin}{\leftmargin}
}

\item 

\textbf{policy evaluation:} the current policy is evaluated, which
includes computing the average cost (\DUrole{ref}{eq-cost}) and the so-called
\emph{value function}
\item 

\textbf{policy improvement:} a single step of optimization with policy
iteration yields a improved policy. Then this policy must be again
evaluated (step 1).\end{list}

The policy evaluation involves solving the equilibrium equation without
the minimization step but with a fixed policy $\mu$ instead:\begin{equation*}
\begin{split}
J_\mu+ \tilde{J}_\mu(x) =
            \underset{w}{\mathbb{E}}  \Big\lbrace  \underbrace{ c(x, \mu(x), w)
                          }_{\text{instant cost}}\\
            +
            \underbrace{ \tilde{J}_\mu(f(x, \mu(x), w))
                         }_{\text{cost of the future}}
             \Big\rbrace   \end{split}
\end{equation*}It can be solved by iterating the right-hand side like for policy
iteration but much faster due to the absence of minimization. In the
end, a few policy improvement iterations are needed to reach
convergence. More details about the value and policy iteration
algorithms can be found in \cite{Bertsekas-2005} textbook for
example. The conditions for the convergence, omitted here, are also
discussed.

\subsection{StoDynProg library description%
  \label{stodynprog-library-description}%
  \label{ss-lib-description}%
}

We have created a small library to \emph{describe}
and \emph{solve} optimal control problems (in discrete time) using the Stochastic
Dynamic Programming (SDP) method. It implements the value iteration and
policy iteration algorithms introduced above.
Source code is available on GitHub \url{https://github.com/pierre-haessig/stodynprog}
under a BSD 2-Clause license.

\subsubsection{Rationale for a library, benefits of Python%
  \label{rationale-for-a-library-benefits-of-python}%
}

Because the SDP algorithms are in fact quite simple (they can be written
with one set of nested for loops) we were once told that they should be
written from scratch for each new problem. However we face in our
research in energy management several optimization problems with slight
structural differences so that code duplication would be
unacceptably high. Thus the motivation to write a unified code that can
handle all our use cases, and hopefully some others’.

StoDynProg is pure Python code built with \texttt{numpy} for
multi-dimensional array computations. We also notably use an external
multidimensional interpolation routine by Pablo Winant (see
\DUrole{ref}{sss-multi-interp} below).

The key aspect of the flexibility of the code is its ability to handle
problems of \emph{arbitrary dimensions} (in particular the state space and
the control space). This impacts particularly the way to iterate over
those variables. Our code makes thus a heavy use of Python tuple
packing/unpacking machinery and \texttt{itertools.product} to iterate on
rectangular grids of arbitrary dimension.

\subsubsection{API description%
  \label{api-description}%
}

StoDynProg provides two main classes: \texttt{SysDescription} and
\texttt{DPSolver}.

\subsubsection{SysDescription%
  \label{sysdescription}%
}

holds the description of the discrete-time dynamic optimization problem.
Typically, a user writes its dynamics function (the Python
implementation of $f$ in (\DUrole{ref}{eq-state-dyn}))
and handles it to a \texttt{SysDescription} instance:\begin{Verbatim}[commandchars=\\\{\},fontsize=\footnotesize]
\PY{k+kn}{from} \PY{n+nn}{stodynprog} \PY{k+kn}{import} \PY{n}{SysDescription}
\PY{c}{\PYZsh{} SysDescription object with proper dimensions}
\PY{c}{\PYZsh{} of state (2), control (1) and perturbation (1)}
\PY{n}{mysys} \PY{o}{=} \PY{n}{SysDescription}\PY{p}{(}\PY{p}{(}\PY{l+m+mi}{2}\PY{p}{,} \PY{l+m+mi}{1}\PY{p}{,} \PY{l+m+mi}{1}\PY{p}{)}\PY{p}{)}

\PY{k}{def} \PY{n+nf}{my\PYZus{}dyn}\PY{p}{(}\PY{n}{x1}\PY{p}{,} \PY{n}{x2}\PY{p}{,} \PY{n}{u}\PY{p}{,} \PY{n}{w}\PY{p}{)}\PY{p}{:}
    \PY{l+s}{\PYZsq{}}\PY{l+s}{dummy dynamics}\PY{l+s}{\PYZsq{}}
    \PY{n}{x1\PYZus{}next} \PY{o}{=} \PY{n}{x1} \PY{o}{+} \PY{n}{u} \PY{o}{+} \PY{n}{w}
    \PY{n}{x2\PYZus{}next} \PY{o}{=} \PY{n}{x2} \PY{o}{+} \PY{n}{x1}
    \PY{k}{return} \PY{p}{(}\PY{n}{x1\PYZus{}next}\PY{p}{,} \PY{n}{x2\PYZus{}next}\PY{p}{)}

\PY{c}{\PYZsh{} assign the dynamics function:}
\PY{n}{mysys}\PY{o}{.}\PY{n}{dyn} \PY{o}{=} \PY{n}{my\PYZus{}dyn}
\end{Verbatim}
We use here a setter/getter approach for the \texttt{dyn} property. The same
is used to describe the cost function ($c$ in (\DUrole{ref}{eq-cost})).
We believe the
property approach enables simplified user code compared to a class
inheritance mechanism. With some inspiration of Enthought \texttt{traits},
the setter has a basic validation mechanism that checks the signature of
the function being assigned (with \texttt{getargspec} from the \texttt{inspect}
module).

\subsubsection{DPSolver%
  \label{dpsolver}%
}

holds parameters that tunes the optimization process, in particular the
discretized grid of the state. Also, it holds the code of the
optimization algorithm in its methods. We illustrate here the creation
of the solver instance attached to the previous system:\begin{Verbatim}[commandchars=\\\{\},fontsize=\footnotesize]
\PY{k+kn}{from} \PY{n+nn}{stodynprog} \PY{k+kn}{import} \PY{n}{DPSolver}
\PY{c}{\PYZsh{} Create the solver for `mysys` system:}
\PY{n}{dpsolv} \PY{o}{=} \PY{n}{DPSolver}\PY{p}{(}\PY{n}{mysys}\PY{p}{)}
\PY{c}{\PYZsh{} state discretization}
\PY{n}{x1\PYZus{}min}\PY{p}{,} \PY{n}{x1\PYZus{}max}\PY{p}{,} \PY{n}{N1} \PY{o}{=} \PY{p}{(}\PY{l+m+mi}{0}\PY{p}{,} \PY{l+m+mf}{2.5}\PY{p}{,} \PY{l+m+mi}{100}\PY{p}{)}
\PY{n}{x2\PYZus{}min}\PY{p}{,} \PY{n}{x2\PYZus{}max}\PY{p}{,} \PY{n}{N2} \PY{o}{=} \PY{p}{(}\PY{o}{\PYZhy{}}\PY{l+m+mi}{15}\PY{p}{,} \PY{l+m+mi}{15}\PY{p}{,} \PY{l+m+mi}{100}\PY{p}{)}
\PY{n}{x\PYZus{}grid} \PY{o}{=} \PY{n}{dpsolv}\PY{o}{.}\PY{n}{discretize\PYZus{}state}\PY{p}{(}\PY{n}{x1\PYZus{}min}\PY{p}{,} \PY{n}{x1\PYZus{}max}\PY{p}{,} \PY{n}{N1}\PY{p}{,}
                                 \PY{n}{x2\PYZus{}min}\PY{p}{,} \PY{n}{x1\PYZus{}max}\PY{p}{,} \PY{n}{N2}\PY{p}{)}
\end{Verbatim}
Once the problem is fully described, the optimization can be launched by
calling \texttt{dpsolv.policy\_iteration} with proper arguments about the
number of iterations.

For more details on StoDynProg API usage, an example problem of
\emph{Inventory Control} is treated step-by-step in the documentation
(created with Sphinx).

\subsubsection{Multidimensional Interpolation Routine%
  \label{multidimensional-interpolation-routine}%
  \label{sss-multi-interp}%
}

StoDynProg makes an intensive use of a multidimensional interpolation
routine that is not available in the “standard scientific Python stack”.
Interpolation is needed because the algorithm manipulates two scalar
functions which are discretized on a grid over the state space: the
value function $\tilde{J}$ and feedback policy $\mu$. Thus,
functions are stored as $n$-d arrays, where $n$ is the
dimension of the state vector ($n=3$ for ocean power smoothing
example). In the course of the algorithm, the value function needs to be
evaluated between grid points, thus the need for interpolation.

\subsubsection{Requirements and Algorithm Selection%
  \label{requirements-and-algorithm-selection}%
}

No “fancy” interpolation method is required so linear interpolation is a
good candidate. Speed is very important because it is called many times.
Also, it should accept vectorized inputs, so that interpolation of
multiple points can be done efficiently in one call.
We assert that the functions will be stored on
a \emph{rectangular grid} which should simplify interpolation computations.
The most stringent requirement is \emph{multidimensionality} (for
$0 \leq n \leq 4$) which rules out most available tools.

We have evaluated 4 routines (details available in an IPython Notebook
within StoDynProg source tree):%
\begin{itemize}

\item 

\texttt{LinearNDInterpolator} class from \texttt{scipy.interpolate}
\item 

\texttt{RectBivariateSpline} class from \texttt{scipy.interpolate}
\item 

\texttt{map\_coordinates} routine from \texttt{scipy.ndimage}
\item 

and \texttt{MultilinearInterpolator} class written by Pablo Winant within
its Dolo project \cite{Winant-2010} for Economic modelling (available on
\url{https://github.com/albop/dolo}).
\end{itemize}

The most interesting in terms of performance and off-the-shelf
availability is \texttt{RectBivariateSpline} which exactly meets our needs
expect for multidimensionality because it’s limited to $n=2$.
\texttt{LinearNDInterpolator} has no dimensionality limitations but works
with unstructured data and so does not leverage the rectangular
structure. Interpolation time was found 4 times longer in 2D, and
unacceptably long in 3D. Then \texttt{map\_coordinates} and
\texttt{MultilinearInterpolator} were found to both satisfy all our
criterions but the latter being consistently 4 times faster (both 2D and
3D case). Finally we also selected \texttt{MultilinearInterpolator} because
it can be instantiated to retain the data once and then called several
time. We find the usage of this object-oriented interface more
convenient than functional interface of \texttt{map\_coordinates}.

\subsection{Results for Searev power smoothing%
  \label{results-for-searev-power-smoothing}%
  \label{ss-results-searev-smooth}%
}
\begin{figure}[]\noindent\makebox[\columnwidth][c]{\includegraphics[width=\columnwidth]{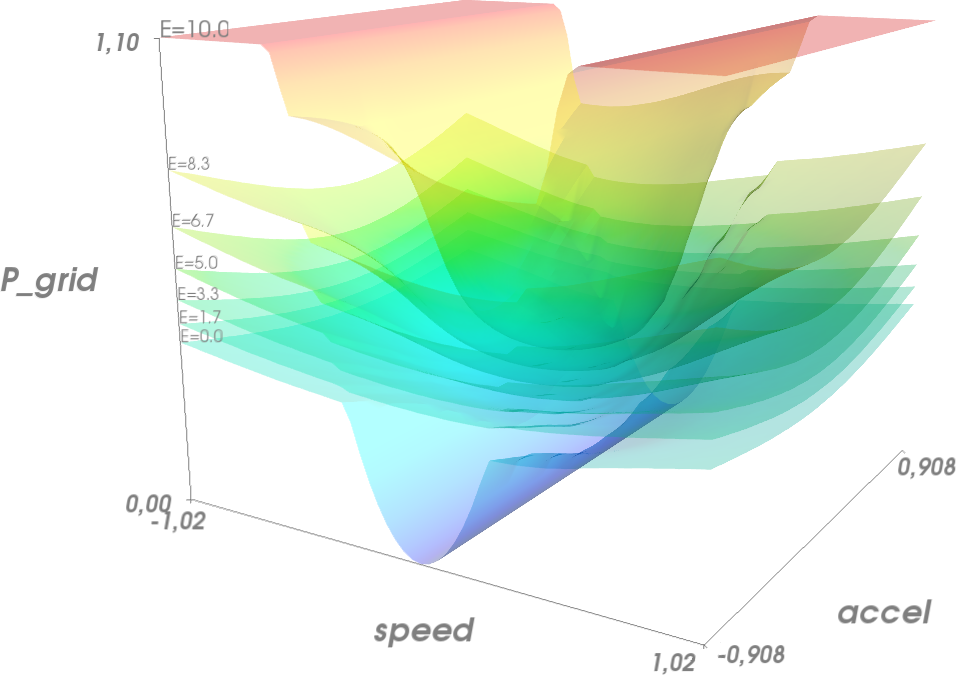}}
\caption{Storage control policy: Power injected to the grid as
a function of speed and acceleration,
for 7 levels of stored energy between empty and full.
\DUrole{label}{fig-P-grid-law}}
\end{figure}\begin{figure*}[]\noindent\makebox[\textwidth][c]{\includegraphics[scale=0.70]{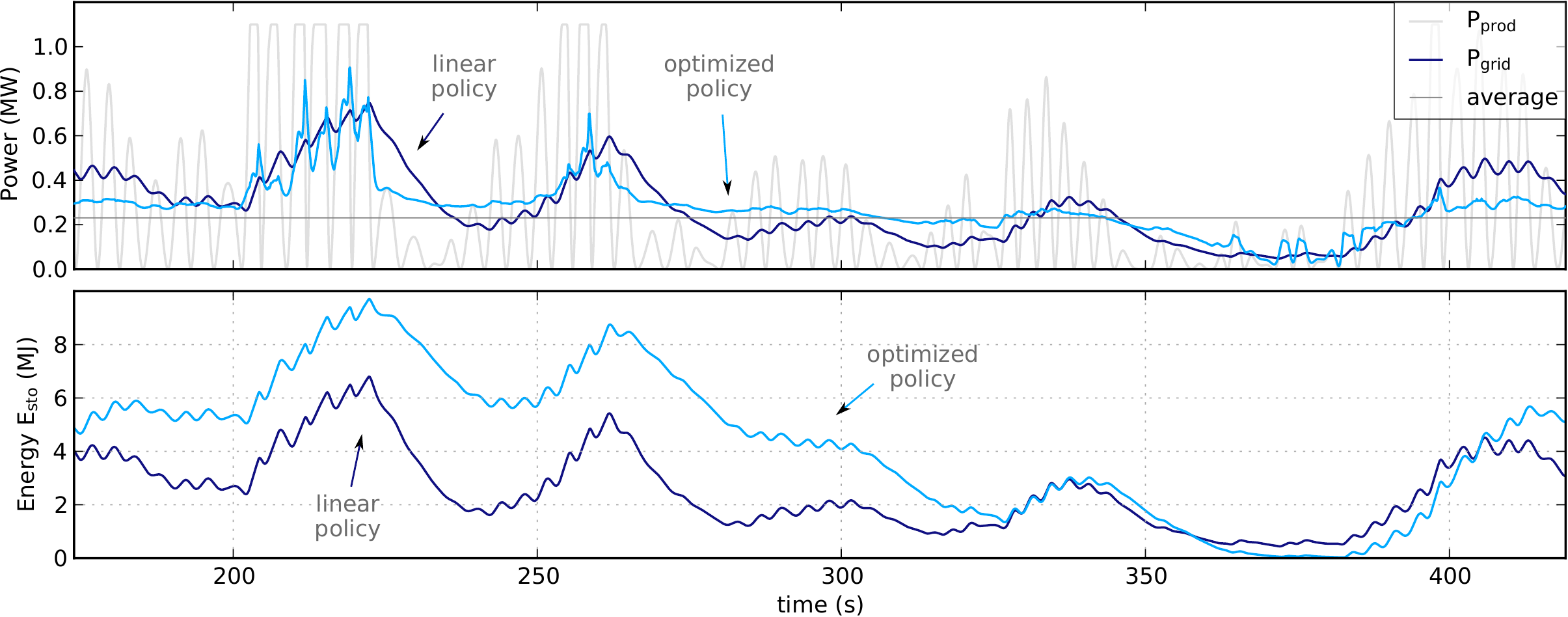}}
\caption{Comparison of the power smoothing behavior between
the \emph{heuristic} (dark blue) and \emph{optimized} (light blue)
storage management policies (storage capacity of 10 MJ).
Stored energy on the bottom panel.
\DUrole{label}{fig-policy-compare}}
\end{figure*}

We have applied the policy iteration algorithm to the SEAREV power smoothing
problem introduced in section~\DUrole{ref}{s-intro-smoothing}.
The algorithm is initialized with the linear storage control policy
(\DUrole{ref}{eq-feedback-lin}).
This heuristic choice is then gradually improved by each policy improvement
step.

\subsubsection{Algorithm parameters%
  \label{algorithm-parameters}%
}

About 5 policy iterations only are needed to converge to an optimal
strategy. In each policy iteration, there is a policy evaluation step
which requires 1000 iterations to converge. This latter number is
dictated by the time constant of the system (1000 steps
$\leftrightarrow$ 100 seconds) and 100 seconds is the time it
takes for the system to “decorrelate”, that is loose memory of its state
(both speed and stored energy).

We also need to decide how to discretize and bound the state space of
the \{SEAREV + storage\} system:%
\begin{itemize}

\item 

for the stored energy $E_{sto}$, bounds are the natural limits
of the storage: $E_{sto} \in [0, E_{rated}]$. A grid of 30
points yields precise enough results.
\item 

for the speed $\Omega$ and the acceleration $A$, there
are no natural bounds so we have chosen to limit the values to
$\pm4$ \emph{standard deviations}. This seems wide enough to include
most observations but not too wide to keep a good enough resolution.
We use grids of 60 points to keep the grid step small enough.
\end{itemize}

This makes a state space grid of
$30\times 60 \times 60 \approx 110\text{k}$ points. Although this
number of points can be handled well by a present desktop computer, this
simple grid size computation illustrates the commonly known weakness of
Dynamic Programming which is the “Curse of Dimensionality”. Indeed, this
size grows exponentially with the number of dimensions of the state so
that for practical purpose state dimension is limited to 3 or 4. This
explains the motivation to search a low order model for the power
production time series in section \DUrole{ref}{s-stoch-model}.

\subsubsection{Algorithm execution time%
  \label{algorithm-execution-time}%
}

With the aforementioned discretization parameters, policy evaluation
takes about 10 s (for the 1000 iterations) while policy improvement
takes 20 s (for one single value iteration step). This makes 30 s in
total for one policy iteration step, which is repeated 5 times.
Therefore, the optimization converges in about 3 minutes. This duration
would grow steeply should the grid be refined.

As a comparison of algorithm efficiency, the use of \emph{value iteration}
would takes much longer than \emph{policy iteration}. Indeed, it needs 1000
iterations, just like policy evaluation (since it is dictated by the
system’s “decorrelation time”) but each iteration involves a costly
optimization of the policy so that it takes 20 s. This makes altogether
5 hours of execution time, i.e. 100 times more than policy iteration!

As possible paths to improve the execution time, we see, at the \emph{code
level}, the use of more/different vectorization patterns although
vectorized computation is already used a lot. Maybe the use of Cython
may speed up unavoidable loops but this may not be worth the loss of
flexibility and the decrease in coding speed. Optimization at the
\emph{algorithm level}, just as demonstrated with “policy vs. value
iteration”, is also worth investigating further. In the end, more use of
Robert Kern’s \texttt{line\_profiler} will be needed to decide the next step.

\subsubsection{Output of the computation%
  \label{output-of-the-computation}%
}

The policy iteration algorithm solves equation (\DUrole{ref}{eq-dp-avg-equil})
and outputs the minimized
cost $J$ and two arrays: function $\tilde{J}$ (transient
cost) and function $\mu$ (optimal policy (\DUrole{ref}{eq-feedback-opt})),
both expressed on the discrete state grid (3d grid).

We focus on $\mu$ which yields the power $P_{grid}$ that
should be injected to the grid for any state of the system.
Figure \DUrole{ref}{fig-P-grid-law} is a Mayavi surface plot which shows
$P_{grid}(\Omega, A)$ for various levels of $E_{sto}$.
Observations of the result are in agreement with what can be expected
from a reasonable storage control:%
\begin{itemize}

\item 

the more energy there is in the storage, the more power should be
injected to the grid (similar to the heuristic control (\DUrole{ref}{eq-feedback-lin})).
\item 

the speed and acceleration of the SEAREV also modulates the injected power,
but to a lesser extent. We may view speed and acceleration as
approximate measurements of the \emph{mechanical energy} of the SEAREV. This
energy could be a hidden influential state variable, in parallel with
the stored energy.
\item 

the injected power is often set between 0.2 and 0.3 MW, that is
\emph{close to the average} power production.
\end{itemize}

Such observations show that the algorithm has \emph{learned} from the SEAREV
behavior to take sharper decisions compared to the heuristic policy it
was initialized with.

\subsubsection{Qualitative analysis of the trajectory%
  \label{qualitative-analysis-of-the-trajectory}%
}

To evaluate the storage control policy, we simulate its effect on the
sample SEAREV data we have (instead of using the state space model used for the
optimization). The only adaptation required for this trajectory
simulation is to transform the \emph{policy array} ($\mu$ known on the
state grid only) into a \emph{policy function} ($\mu$ evaluable on the
whole state space). This is achieved using the same n-dimensional
interpolation routine used in the algorithm.

A simulated trajectory is provided on figure
\DUrole{ref}{fig-policy-compare} to compare the effect of the optimized
policy with the heuristic linear policy (\DUrole{ref}{eq-feedback-lin}).
As previously said, the
storage capacity is fixed at $E_{rated}=10\text{ MJ}$ or about 9
seconds of charge/discharge at the rated power.

Positive aspect, the optimized policy yields an output power that is
generally closer to the average (thin gray line) than the linear policy.
This better smoothing of the “peaks and valleys” of the production is
achieved by a better usage of the available storage capacity. Indeed,
the linear policy generally under-uses the higher levels of energy.

As a slight negative aspect, the optimized policy yields a “spiky”
output power in the situations of high production (200–220 s). In this
situation, the output seems worse that the linear policy. We connect
this underperformance to the linear model (\DUrole{ref}{eq-ss-ar2})
used to represent the SEAREV dynamics.
The linearity holds well for small movements but not when the
speed is high and the pendulum motion gets very abrupt (acceleration
high above 4 standard deviations which contradicts the Gaussian
distribution assumption). Since the control optimization is based on the
linear model, the resulting control law cannot appropriately manage
these non-linear situations. Only an upgraded model would genuinely
solve this problem but we don’t have yet an appropriate low-order
non-linear model of the SEAREV. One quick workaround to reduce the power peaks
is to shave the acceleration measurements (not demonstrated here).

\subsubsection{Quantitative assessment%
  \label{quantitative-assessment}%
}

We now numerically check that the optimized policy brings a true
enhancement over the linear policy. We simulate the storage with the
three 1000 s long samples we have and compute the power variability
criterion\DUfootnotemark{id12}{id14}{2} for each.

Figure \DUrole{ref}{fig-policy-assess} shows the standard deviation for each
sample in three situations: without storage (which yields the natural
standard deviation of the SEAREV production), with a storage controlled by the
linear policy and finally the same storage controlled by the optimized
policy. Sample \texttt{Em\_1.txt} was used to fit the state space model
(\DUrole{ref}{eq-ss-ar2})
but we don’t think this should introduce a too big bias because of the low
model order.

Beyond the intersample variability, we can see the consistent
improvement brought by the optimized law. Compared to the linear policy,
the standard deviation of the injected power is reduced by about 20~\%
(27~\%, 16~\%, 22~\% for each sample respectively). We can conclude that
the variability of the injected power is indeed reduced by using the
Dynamic Programming.\begin{figure}[]\noindent\makebox[\columnwidth][c]{\includegraphics[scale=0.60]{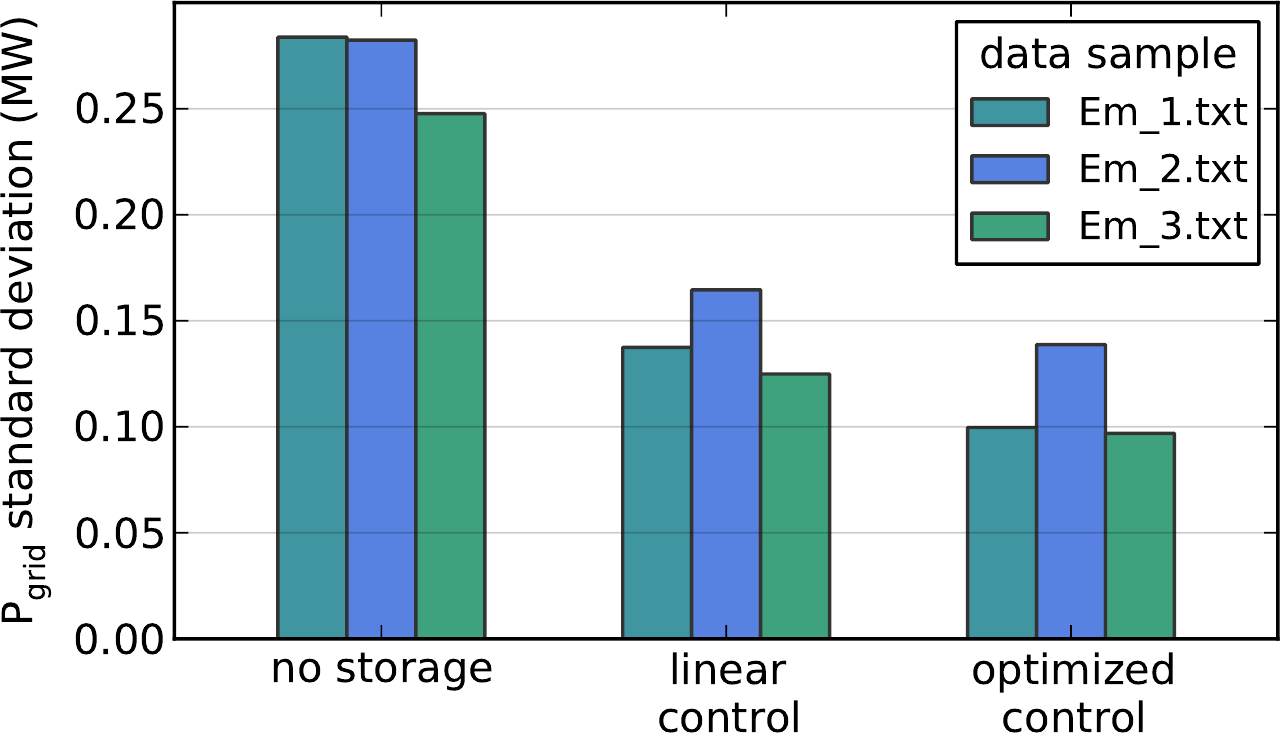}}
\caption{Effect of optimizing the storage control on three SEAREV
production time series.
Standard deviation compared to the heuristic linear control case
is reduced by about 20 \%.
\DUrole{label}{fig-policy-assess}}
\end{figure}

Because the RMS deviation criterion used in this article is not directly limited
or penalized in current grid codes, there is no financial criterion
to decide whether the observed deviations are acceptable or not.
Therefore we cannot conclude if the \textasciitilde{}20\% reduction of the variability brought
by optimal control is valuable.
Nevertheless, there exists criterions like the \textquotedbl{}flicker\textquotedbl{} which are used
in grid codes to set standards of power quality.
Flicker, which is way more complicated to than an additive criterion like
(\DUrole{ref}{eq-cost}) could be used to put an economic value on a control strategy.
This is the subject of ongoing research.

\section{Conclusion%
  \label{conclusion}%
}

With the use of standard Python modules for scientific computing, we
have created StoDynProg, a small library to solve Dynamic Optimization
problems using Stochastic Dynamic Programming.

We have described the mathematical and coding steps required to apply the
SDP method on an example problem of realistic complexity: smoothing
the output power of the SEAREV Wave Energy Converter. With its generic
interface, StoDynProg should be applicable to other Optimal Control
problems arising in Electrical Engineering, Mechanical Engineering or
even Life Sciences. The only requirement is an appropriate mathematical
structure (Markovian model), with the “Curse of Dimensionality”
requiring a state space of low dimension.

Further improvements on this library should include a better source tree
organization (make a proper package) and an improved test coverage.
%
\DUfootnotetext{id13}{id4}{1}{
In the special case of a linear dynamics and a quadratic cost (“LQ
control” ), the optimal feedback is actually a \emph{linear} function.
Because of the state constraint
$0 \leq E_{sto} \leq E_{rated}$, the storage control problem
falls outside this classical case.}
\DUfootnotetext{id14}{id12}{2}{
Instead of using the exact optimization cost (\DUrole{ref}{eq-cost})
(average quadratic power in MW\textsuperscript{2}),
we actually compute the standard deviation (in MW).
It is mathematically related to the quadratic power and we find it more
readable.}

\end{document}